# $^6$Li-loaded liquid scintillators produced by direct dissolution of compounds in diisopropylnaphthalene (DIPN)

N. P. Zaitseva*, M. L. Carman, M. J. Ford, A. M. Glenn, Ch. Roca, S. Durham, F. Sutanto, S. A. Dazeley, N. S. Bowden

$^a$Lawrence Livermore National Laboratory, 7000 East Avenue, Livermore, CA94550, USA

**Abstract**

The paper describes preparation of $^6$Li-loaded liquid scintillators by methods involving direct dissolution of $^6$Li salts in the commercial diisopropylnaphthalene (DIPN) solvent, without the formation of water-in-oil emulsions. Methods include incorporation of $^6$Li that, unlike previously reported formulations, does not require additions of water or a strong acid such as hydrochloric acid (HCl). Results of the conducted experiments show that dissolution of aromatic and aliphatic $^6$Li salts in DIPN can be easily achieved at 0.1- 0.3% by weight of atomic $^6$Li, using small additions of waterless surfactants, or mild carboxylic acids. An alternative way suggests incorporation of $^6$Li as a part of a surfactant molecule that can be dissolved in DIPN without any solubilizing additions. Proposed methods enable preparation of efficient $^6$Li-loaded liquid scintillators that, at a large scale of 50 cm, exhibit good pulse shape discrimination (PSD) properties combined with up to 107% of light output and up to 115% of the attenuation length measured relative to standard undoped EJ-309 liquid scintillator.

---

\* Corresponding author.

E-mail address: zaitseva1@llnl.gov (Natalia Zaitseva)

1. **Introduction**

Organic liquid scintillators have been known for decades as excellent materials for fast neutron detection. Efficient fluorescence combined with high hydrogen content determines their pulse shape discrimination (PSD) properties that enable the separation of neutron signatures from the gamma ray background [1, 2]. Since the composition of organic scintillators is comprised of mostly hydrocarbons, traditional liquid scintillators can be used only for detection of fast neutrons, leaving undetected the large fraction of low-energy (thermal) neutrons that do not generate enough light via (*n,p*) scattering. Loading of organic scintillators with $^6$Li nuclei is known as one of the ways to achieve additional sensitivity to thermal neutrons. $^6$Li possesses many desirable qualities as a neutron capture agent. Among its important advantages are a reasonable capture cross section, relatively high photon yield from charged particles, and absence of gamma-rays in the final products resulting from a thermal neutron capture reaction $^6$Li + n$^o$ = $^3$H + α + 4.8 MeV. The principal drawback, however, is very low or no solubility of highly polar $^6$Li-containing compounds in non-polar aromatic solvents needed for efficient scintillation. Early attempts [3-5] to incorporate $^6$Li into a liquid scintillator (LS) by simple dissolution of lithium salts of carboxylic acid did not lead to production of stable compositions containing $^6$Li at concentrations required for any practical use. More success has been achieved with a different approach, in which a surfactant added to a LS allows for the loading of an aqueous $^6$LiCl solution to produce a stable microemulsion [6-8]. As reported in recent publications [9,10], a formulation based on this approach was used to produce a large volume of efficient, stable scintillator for the antineutrino detector of the Precision Reactor Oscillation and Spectrum (PROSPECT) experiment at Oak Ridge National Laboratory (ORNL) [11].

The reported generic formula of the PROSPECT LS consists of a nonionic surfactant and a 9.98 mol/L aqueous solution of LiCl containing 95% enriched $^6$Li at about 0.08% by mass fraction. The aqueous phase is dispersed into an organic phase that is comprised of 2,5-diphenyloxazole (PPO) and 1,4-bis(2-methylstyryl) benzene (bis-MSB) dissolved in diisopropylnaphthalene (DIPN) solvent, which is also used as a low toxicity and high flash point base for EJ-309 liquid scintillator [12]. According to the literature reports [9], this formulation satisfies the current acceptance criteria in absorbance, light yield, and PSD capabilities for deployment in the PROSPECT antineutrino detector at ORNL. Future developments, however, may bring new requirements for adjustments and improvement of scintillation performance and physical properties of detector materials. This stimulates continuation of a broader investigation into $^6$Li-containing scintillators that includes large-scale plastics [13-15] and advanced LSs.

The goal of the experiments described in this paper was to explore alternative methods for incorporation of $^6$Li into aromatic solvents. Like in the PROSPECT LS, the studies were conducted using only DIPN solvent to ensure an end product suitable for use and transport in large volumes. However, no water or hydrochloric acid were added to these preparations, meaning that all homogeneous solutions were obtained by direct dissolution of ingredients without use of water as solubilizing addition. New formulations were characterized to determine the effects leading to quenching of light output (LO) and PSD and to select compositions that may be promising for potential applications like neutron and antineutrino detection that would benefit from enhanced scintillation performance.

2. **Experimental**

EJ-309 and DIPN (mixture of isomers) are standard commercial products that can be purchased from Eljen Technology. Most scintillation dyes, such as Exalite E404 (1,4-bis(9,9-diethyl-7-(*tert*-pentyl)-9*H*-fluoren-2-yl)benzene), bis-MSB (1,4-bis(2-methylstyryl)benzene), C460 (7-diethylamino-4-methylcoumarin), and BBQ (4,4'''-bis[(2-butyloctyl)oxy]-1,1':4',1'':4'',1'''-quaterphenyl) were purchased from Exciton/Luxottica. PPO (2,5-diphenyloxazole), DPA (9,10-diphenylanthracene), BP (biphenyl) and Tx-100 (Triton X-100) were produced by Sigma-Aldrich. $^6$Li-salts used for preparation of liquid scintillators were synthesized using reactions between $^6$Li$_2$CO$_3$ and carboxylic acids, among which IBA (isobutyric acid) and IVA (isovaleric acid) were used as received from Sigma Aldrich, while aromatic 3-PSA (3-phenylsalicylic acid, TCI America) was purified by recrystallization from methanol. $^6$Li$_2$CO$_3$ was obtained from the National Isotope Development Center at the Oak Ridge National Laboratory. $^6$Li-salts were synthesized by reacting excess amounts (1.05 mol equivalent acid:1 mol equivalent base) of corresponding acids with $^6$Li$_2$CO$_3$ in methanol. The resulting crude materials were either washed with anhydrous diethyl ether to remove excess acids or precipitated using excess cold acetone. The salts were dried on a Schlenk line at 90°C under nitrogen for four hours to remove residual solvents. $^6$LiDOSS ($^6$Li-1,4-bis(2-ethylhexoxy)-1,4-dioxobutane-2-sulfonate) was prepared by first dispersing an ion exchange resin in a solution of analogous sodium salt NaDOSS (commonly known as dioctyl sulfosuccinate sodium, or Aerosol OT), purchased from Sigma Aldrich. The conjugate acid of NaDOSS was then filtered and reacted with $^6$Li$_2$CO$_3$ (Scheme 1). Alternatively, NaDOSS was added to methanolic HCl to precipitate NaCl, obtaining the organic acid. $^6$Li$_2$CO$_3$ was added in excess and mixed at 90°C overnight to conduct the acid-base reaction that forms the final $^6$LiDOSS (Scheme 1).

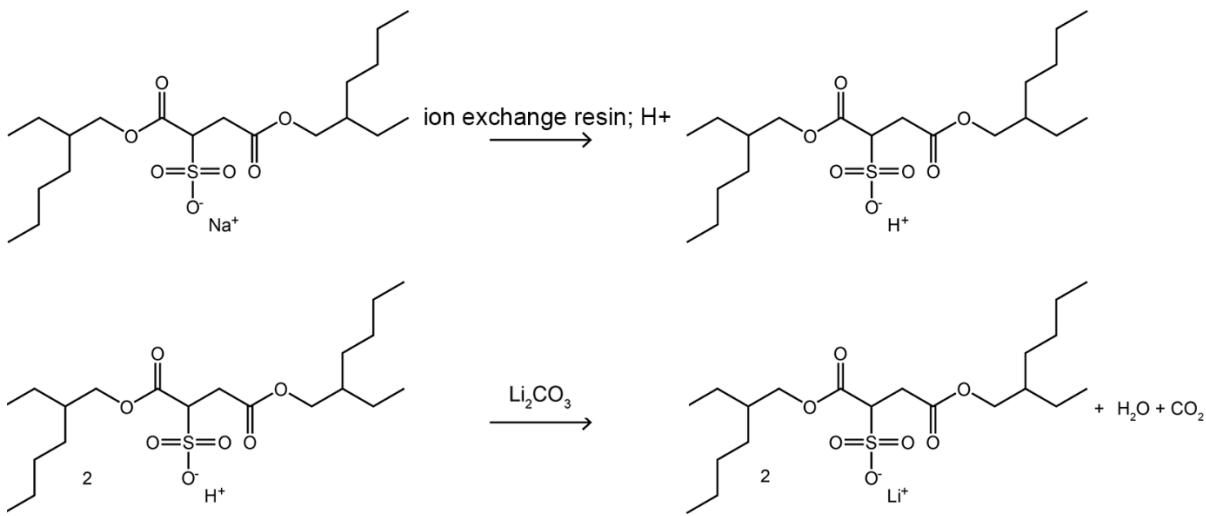

*Scheme 1. Synthesis of LiDOSS*

The excess $^6Li_2CO_3$ was removed by dissolving the $^6$LiDOSS product in toluene and filtering. The product in toluene was dried over magnesium sulphate and filtered. The product was isolated by removing toluene by evaporation. Most scintillator compositions were prepared with 0.1 wt. % of atomic $^6$Li, although higher $^6$Li concentrations were also tested with some efficient formulations. Note that here and below, the concentration of individual components is defined in weight (wt.) % relative to the total weight of the sample.

For liquid preparations, solid constituents weighed in vials or glass jars at ambient condition were transferred into a nitrogen-filled glove box, where liquid components (EJ-309, DIPN, Tx-100, or IBA) were added for dissolution. Solutions were left to stir overnight to reach equilibrium, then transferred into quartz cuvettes, sealed, and removed from the glovebox. Unless indicated otherwise, measurements were made in small cylindrical cuvettes of Ø2 cm × 1 cm in dimensions. The most efficient compositions were additionally measured in 2" × 2" (Ø5 cm × 5 cm) quartz cuvettes, for approximate evaluation of attenuation that was further verified at larger scale.

Photoluminescence (PL) spectra were measured using a Horiba Jobin Yvon Fluoromax-4 spectrometer equipped with a 450 W Xe lamp in reflectance mode. In small-scale experiments, scintillation responses were measured by wrapping the cuvettes in Teflon tape leaving open one flat window, which was coupled using optical grease to a Hamamatsu R6231-100-SEL photomultiplier tube (PMT). Signals collected from the PMT were recorded using a 14-bit high resolution CompuScope 14200 waveform digitizer at a sampling rate of 200 MS/s. LO was calibrated using 667 keV $^{137}$Cs gamma-rays by determining the 50% position of the maximum value of the Compton edge from the pulse integral histogram (i.e. LO in ADC units). This feature is taken as 500 keV. Comparison of the LO for different formulations was made relative to a small (Ø2 cm x 1 cm) EJ-309 standard sample, the LO of which was taken as unity. Neutron/gamma PSD distributions were obtained using a $^{252}$Cf source moderated by 5 cm of lead and one inch (2.54 cm) thick high-density polyethylene. Resulting waveforms were integrated over two-time integrals corresponding to the total charge ($Q_{total}$) and the delayed component of the signal ($Q_{tail}$), respectively. The end point of the intervals was the same for $Q_{tail}$ and $Q_{total}$. The ratio $Q_{tail}/Q_{total}$ indicated whether an event was likely produced by a neutron (larger ratio due a greater fraction of the delayed light in the pulse) or a gamma-ray (smaller ratio). Quantification of the PSD was accomplished by calculating standard figures of merit (FoM)

$$FoM = \frac{\langle n \rangle - \langle \gamma \rangle}{FWHM_n + FWHM_\gamma}$$

where <n> and <γ> are the centroids of the neutron or gamma distributions, and $FWHM_n$ and $FWHM_\gamma$ are the full widths at half maximum of the neutron and gamma-ray distributions, respectively. The beginning and end time of $Q_{tail}$ were varied to find the value that maximizes the FoM for the energy range of interest. A typical $Q_{tail}$ interval is ~40ns to ~750 ns, where 0 ns is the pulse peak; however, the optimal window varied with scintillator composition. For liquids without $^6$Li, fast neutron/gamma-ray PSD was evaluated by calculation of FoM over the energy range 450 – 510 keV$_{ee}$. With $^6$Li-containing liquids, FoM was obtained by fitting the distributions over the energy region ±2.5σ around the centroid of the thermal neutron spot to a Gaussian function. It should be noted that because of the random effects of oxygen quenching or optical coupling between PMT and quartz cuvettes, LO and PSD of samples, that were the same size and composition but prepared in different batches (each composition was prepared and measured in 2-4 individual samples) might show variations within 3%. Based on greater than thirty long-term measurements of a 10g EJ-200 sample, which showed no systematic trend, we estimate a systematic standard deviation uncertainty of 2.8% for LO measurements due to effects such as mounting and temperature variations. We note that the measurements in these experiments were performed over a much shorter period, so we consider this a conservative estimate. Since FoM varies approximately as the square root of light output, the uncertainty for FoM may be less, but we conservatively apply the same 2.8% uncorrelated systematic uncertainty. The maximum LO deviation of ~3% observed among scintillators produced with the same composition provided no indication that manufacturing reproducibility contributed significant additional uncertainty.

The effective attenuation length, relative LO and PSD FoM of the most efficient formulations were measured at larger scale using cells constructed by bonding of 0.64 cm thick panels of UVT acrylic with outer dimension 5.5cm x 5.5 cm x 50 cm. This length and aspect ratio was chosen to be similar to elongated cells used in large, segmented liquid scintillator detectors [9, 16, 17], where light attenuation and transport have an important influence on detector performance. The acrylic cells were wrapped in DF2000MA multi-layer specular reflector and optically coupled to Hamamatsu 2" H11284-100 PMT assemblies at either end. Readout from the PMTs was performed using a CAEN V1725S 14-bit digitizer. Measurement of the effective attenuation length and relative LO was performed using an automatized linear stage transporting a $^{22}$Na source longitudinally across the liquid cells. The source was held inside a 5 cm thick lead collimator to localize energy depositions. Linear stage and data acquisition system were configured to take 20 equidistant $^{22}$Na measurements along the length of the cell. The primary observable was the Compton edge of the 1274 keV gamma ray emitted by $^{22}$Na. The relative LO of each formulation was obtained from comparing the geometric mean of both PMT pulse amplitudes, which approximately corrects for light transport attenuation as a function of longitudinal event position [9, 16, 17]. The effective attenuation length, a parameter that incorporates effects due to the light scatter and absorption properties of the LS, as well as the light transport properties of the reflectorized 50 cm acrylic cell, was calculated from a linear fit to the ratio of PMT pulse amplitudes as a function of source position using the equation ln $S_1/S_2$ = $(-2/\lambda_{eff})\,x + b$. Here $S_1 = C\,exp - x/\lambda_{eff}$ and $S_2 = C\,exp - (x - L)/\lambda_{eff}$ are exponentials corresponding to the PMT's signals, with $C$ being an arbitrary constant proportional to the total light output and $x$ goes from 0 to L representing the position of the source during the scan.

PSD of the large-scale samples was characterized using a similar setup. In this case, the linear stage was removed, and the gamma source was replaced by a $^{252}$Cf source, which was positioned longitudinally at the center of the cells and shielded by 5 cm of lead. The calculation of the FoM was performed following the same prescription described above for the smaller liquid samples.

### 3. Results and discussions

*3.1. Small-scale experiments with (Ø2 cm x 1 cm) samples*

*3.1.1. $^6$Li-loading of EJ-309*

Addition of lithium to EJ-309, which is commercially available and has high scintillation LO and excellent PSD (Figure 1A), would be a simple way to obtain a $^6$Li-loaded LS. However, since lithium salts are practically insoluble in pure aromatic solvents, incorporation of $^6$Li into organic scintillators always requires the use of some solubilizing compounds. With the PROSPECT liquids, dissolution of $^6$Li was achieved using surfactants that are long molecules with hydrophilic and hydrophobic ends. Incorporation of water-dissolved $^6$LiCl into EJ-309 liquid becomes possible due to the formation of small micelles, where the hydrophilic ends of the surfactant stabilize the polar water droplets, and the opposite hydrophobic ends enable solubility of these micelles in nonpolar aromatic medium. A similar approach can be used in the absence of water, by direct dissolution of some $^6$Li-salts of aromatic acids combined with common surfactants, such as Tx-100. Our experiments showed that the addition of 3-6 wt.% of Tx-100 allows for easy dissolution of several $^6$Li-salts of different salicylic acids, such as $^6$Li salicylate, $^6$Li-3-methylsalicylate, or $^6$Li-3,5-tertbutylsalicylate. Some of these salts, like $^6$LiPSA ($^6$Li salt of 3-phenylsalicylic acid) reported previously [18], may be slightly soluble in aromatics, but addition of Tx-100 substantially increases solubility of the prepared mixtures that, at 0.1 wt.% of atomic $^6$Li, remain stable to precipitation after >1 year of observation. As illustrated by Figure 1 A and B, PSD distributions that can be obtained in LSs containing aromatic $^6$Li salts (e.g., $^6$LiPSA) produce clear scintillation signals from gamma rays, fast neutrons, and thermal neutrons, only slightly decreasing the fast neutron/gamma separation in comparison to EJ-309. The disadvantage of Tx-100 is that it still works via the micelle-formation mechanism that may lead to long-term instability and potential degradation of scintillation performance. Another problem is that this surfactant can be used as a solubilizing agent only with a very limited group of $^6$Li salts, specifically, Tx-100 was found to be effective only with a few salts of salicylic acids, which require complex purification and synthesis procedures. And finally, $^6$Li-salicylates are fluorescent compounds that emit at relatively long wavelengths (~430 nm). Thus, they may function as inefficient secondary dyes, quenching the final LO of the traditional scintillation systems emitting at shorter wavelengths of 400-420 nm [14].

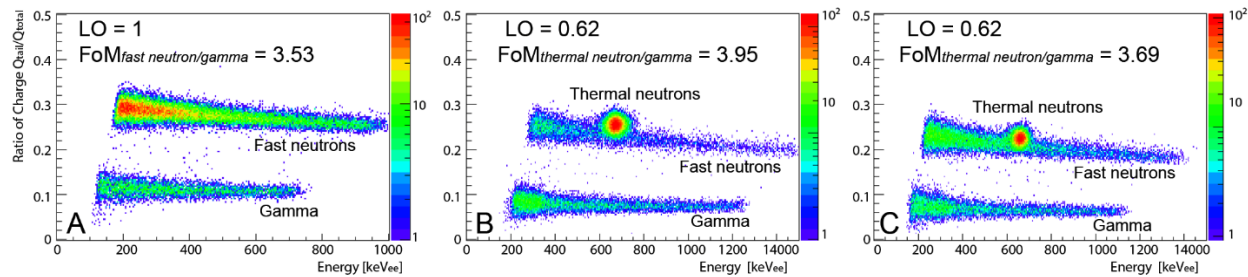

*Figure 1. Typical PSD distributions measured with EJ-309-based liquid scintillators: A – Initial EJ-309; B – EJ-309 loaded with $^6$LiPSA + 3% Tx-100; C – EJ-309 loaded with $^6$LiIVA +10% IBA. The concentration of atomic $^6$Li in both B and C is about 0.1 wt.%.*

A simpler and more practical way to incorporate non-aromatic (e.g., aliphatic) $^6$Li salts can be achieved through direct dissolution, using small additions of carboxylic acids as solubilizing agents. This method has been applied to plastic scintillators loaded with short chain $^6$Li salts of butyric, methacrylic, and pivalic acids [14, 19, 20]. Experimental studies showed that most of these aliphatic salts can be dissolved in EJ-309 using 5-10 wt. % additions of isobutyric (IBA) or methacrylic (MAA) acids. Figure 1C presents an example of PSD distributions obtained with the most soluble $^6$Li salt of isovaleric acid ($^6$LiIVA), dissolved in a 90:10 ratio mixture of EJ-309 and IBA that exhibits a narrow thermal neutron peak positioned at relatively high energy around 600-650 keV$_{ee}$ (Figure 2A). Despite the good discrimination properties, however, suggested methods for aromatic or aliphatic $^6Li$ dissolution suffer from the same problem, which is a substantial loss of the LO in $^6Li$-loaded compositions compared to the initial, non-loaded EJ-309 (Figure 2B). As follows from measurements made with many samples, a typical decrease of the LO upon loading EJ-309 with different $^6Li$-salts, results in ~35-40% LO loss, which is close to the ~30% reported for the PROSPECT LS [9]. Considering that in both cases of the PROSPECT and the formulations described above, EJ-309 was used as a base, this similarity raises a question of whether the low LO is an inherent effect due to the addition of $^6Li$ or whether compositions alternative to EJ-309 can be found to prevent or to diminish the LO loss in $^6Li$-loaded LSs.

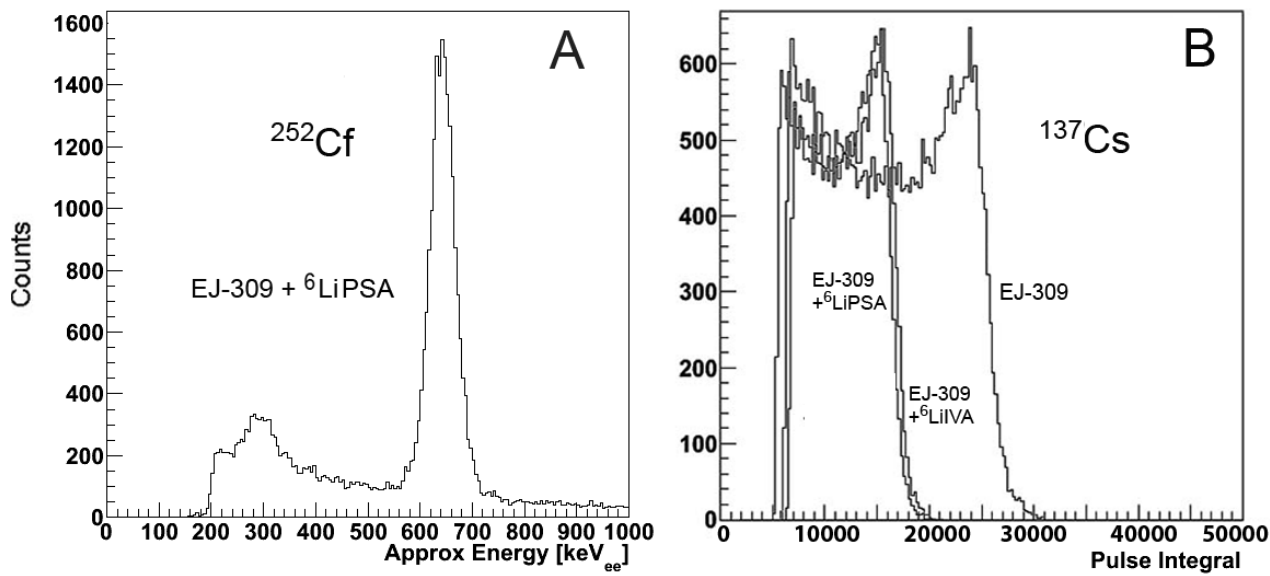

*Figure 2. A – $^{252}Cf$ spectrum of a $^6LiIVA$-loaded EJ-309 liquid showing thermal neutron peak response at 630KeV$_{ee}$; B – Superposition of the $^{137}Cs$ Compton edges showing ~40% LO loss in EJ-309 loaded by $^6LiPSA$ and $^6LiIVA$ (0.1 wt.% $^6Li$), in comparison with the same size unloaded EJ-309.*

*3.1.2. DIPN solutions containing different scintillating dyes (no $^6Li$ added)*

LSs provide great opportunities for studying compositional effects because efficient PSD in liquid solutions can be obtained at very low concentrations achievable with many low-solubility dyes. The wide range of LSs that can be formulated to study PSD can be contrasted with the range of compositions that allow for PSD in plastic scintillators, which rely on high concentration of dyes (up to ~ 30 wt. %) [21]. High mobility of molecules in liquids provides efficient interaction of singlet and triplet excited states where, depending on the individual fluorescent dyes, the maximum LO and efficient PSD can be achieved at concentrations of 0.1-0.5 wt. % (Figure 3). Historically, LSs were made using solvents with low flashpoints, like *p*-xylene or pseudocumene, and were prepared using pairs of primary dyes (e.g., PPO, *p*-terphenyl) and secondary dyes (wavelength shifters, e.g., POPOP or bis-MSB). Even though efficient PSD could be achieved at low concentration of single dyes, systems that contain two dyes were needed to facilitate efficient energy transfer from the solvents emitting at very short wavelengths (~290 nm) to the wavelength shifters absorbing and emitting at much longer wavelengths (e.g., peak absorption of POPOP is ~360 nm with emission at ~420 nm).

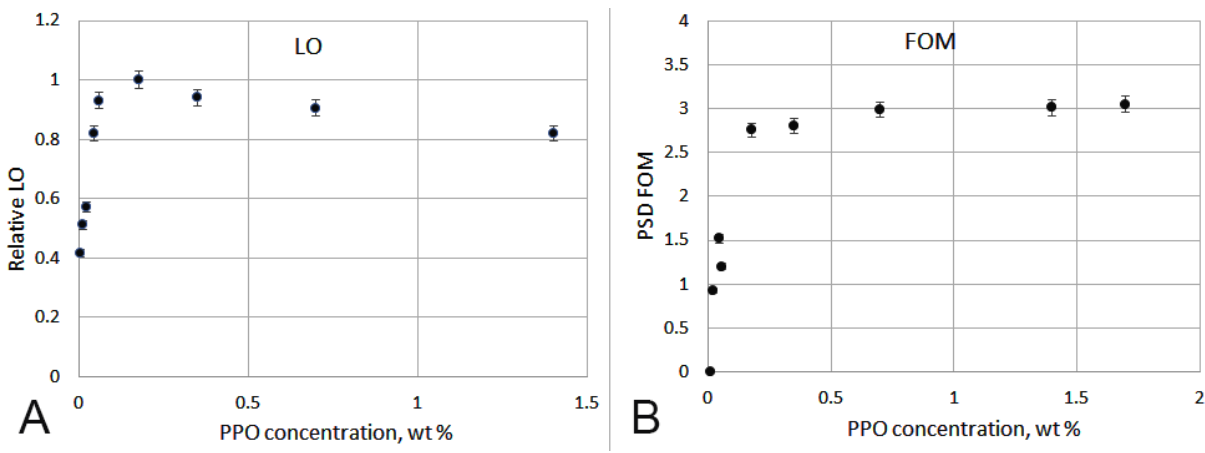

*Figure 3. A – Typical dependence of LO vs concentration measured with PPO dissolved in p-xylene with the maximum LO at ~ 0.2 wt.% taken as a unity; B – Corresponding FoM dependence showing that in liquids the low dye concentration of 0.2-0.5 wt.% is sufficient to achieve maximum PSD.*

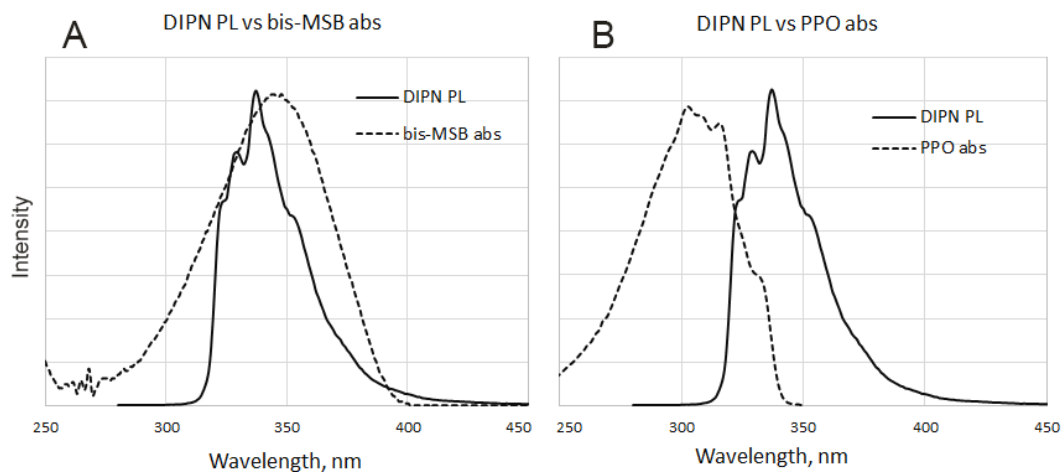

*Figure 4. Superposition of the spectra showing (A) – Perfect overlap between DIPN PL emission and bis-MSB absorption that provides efficient excitation transfer of excitation energy from DIPN directly to bis-MSB; B – Much poorer overlap between DIPN PL and PPO absorption diminishes the role of the primary dye in the final scintillation efficiency produced in DIPN solutions mainly by bis-MSB.*

The advantage of DIPN is the emission spectrum that is shifted to longer wavelengths matching the absorption spectra of most dyes traditionally used as wavelength shifters. For example, in an LS containing PPO and bis-MSB (Figure 4), there is good spectral overlap between DIPN PL and bis-MSB absorption, which enables efficient energy transfer from DIPN to bis-MSB. The direct energy transfer from the solvent to the secondary dye diminishes the role of PPO that, due to the poorer overlap of the absorption of PPO with the emission of DIPN, may not participate in singlet energy transfer. These spectral relationships are reflected in the LO and PSD parameters measured with liquid scintillators prepared by addition of only one secondary dyes (bis-MSB, C460, E404, and BBQ), or the same dyes combined in pairs with some primary dyes (PPO or BP) (Table 1).

|  | EJ-309 | PPO + bis-MSB | | | | PPO + C460 | | PPO (BP) + E404 | | | BBQ |
|---|---|---|---|---|---|---|---|---|---|---|---|
| Primary dye % |  | 0.2 | 0 | 0.2 | 0.2 | 0 | 0.2 | 0 | 0.2 (PPO) | 0.2 (BP) | 0 |
| Secondary dye % |  | 0 | 0.2 | 0.2 | 0.05 | 0.2 | 0.2 | 0.5 | 0.5 | 0.5 | 0.5 |
| Relative LO | 1 | 0.90 | 1.00 | 0.98 | 1.00 | 1.00 | 0.98 | 1.04 | 1.04 | 0.97 | 1.09 |
| FoM | 3.53 | 3.57 | 3.24 | 3.20 | 3.46 | 3.59 | 3.64 | 3.71 | 3.65 | 3.48 | 3.72 |
13

*Table 1. Relative LO and PSD measured with DIPN-based liquid scintillators containing single dyes (bis-MSB, C460, E404, and BBQ) alone or in combination with the primary dye (PPO or BP). Results show that scintillation parameters close to those of EJ-309 can be obtained with single-dye systems, independent of the presence or absence of the primary dye. The systematic uncertainty to LO and FoM values is 2.8%.*

As follows from the comparison, independent of the presence or absence of primary dyes, LSs exhibit high LO and PSD that is close to those of EJ-309. The performance seems to be determined mainly by the properties of the secondary dyes since DIPN allows for efficient excitation transfer to these dyes. In some cases, the optimal performance of LSs can be achieved by addition of a primary dye or slight tuning of concentrations, as may be done in traditional EJ-309. However, the overall results of Table 1 show that, with the DIPN solvent, efficient LSs may be made with single dyes, some of which, like E404 or BBQ, may produce LO and PSD superior to those of EJ-309. The practicality of such an approach requires additional considerations related to the cost, availability, and attenuation at increasing scale. As for the subject of the current discussion, efficient DIPN systems prepared by addition of single fluors provide convenient testbeds for more detailed consideration of the effects produced by $^6$Li additions on the performance of individual dyes in LSs.

### 3.1.3. $^6$Li incorporation into single-dye systems

Figure 5A shows the LO change produced by incorporation of $^6$Li into compositions of EJ-309 and other DIPN-based LSs containing different single dyes. For these experiments, $^6$Li- loaded samples were prepared by dissolution of 1.79 wt. % of $^6$LiIVA (corresponding to ~ 0.1wt. % of $^6$Li) in 90:10 or 95:5-ratio mixtures of EJ-309 or DIPN with IBA. Analysis of the results shows that, among all tested formulations, addition of $^6$Li produces the highest impact on the PPO-DIPN solution, in which LO decreases by ~ 45% upon addition of $^6$LiIVA in combination with 10% IBA. A slightly smaller but similar LO loss is also observed for $^6$Li and 10% IBA loaded into EJ-309 that may also use PPO as a primary dye. With other, non-PPO solutions, the effect of $^6$Li-loading is milder. Moreover, with most of these compositions, the dissolution of the same amounts of $^6$LiVA can be fully achieved with smaller additions of the solubilizing acid, down to 95:5 DIPN:IBA ratio (exception is EJ-309 that shows signs of solid precipitation in a few days after the preparation). A visible rise of the LO at the decreasing IBA concentration indicates that, in addition to lithium, small amounts of acids may be a strong factor affecting the LO in organic scintillators. Since neither $^6$LiIVA nor IBA absorb in the optical range of any of the dyes' emission, the negative effect of these compounds could be attributed to a possible formation of some light-absorbing complexes that form between the additives and the aromatic constituents of the LSs. With this suggestion, the reactivity of PPO related to the presence of oxygen and nitrogen in its composition may explain the larger drop of the LO in PPO-DIPN solution relative to other studied LSs. The sensitivity of PPO to these additives relative to other dyes can be highlighted by examining non-PPO DIPN LSs that exhibit less dramatic LO loss with a smaller dependence on the acid concentration. This trend is especially evident for the solutions containing pure hydrocarbons (DPA, bis-MSB, and E404), which may be less reactive with the polar compounds.

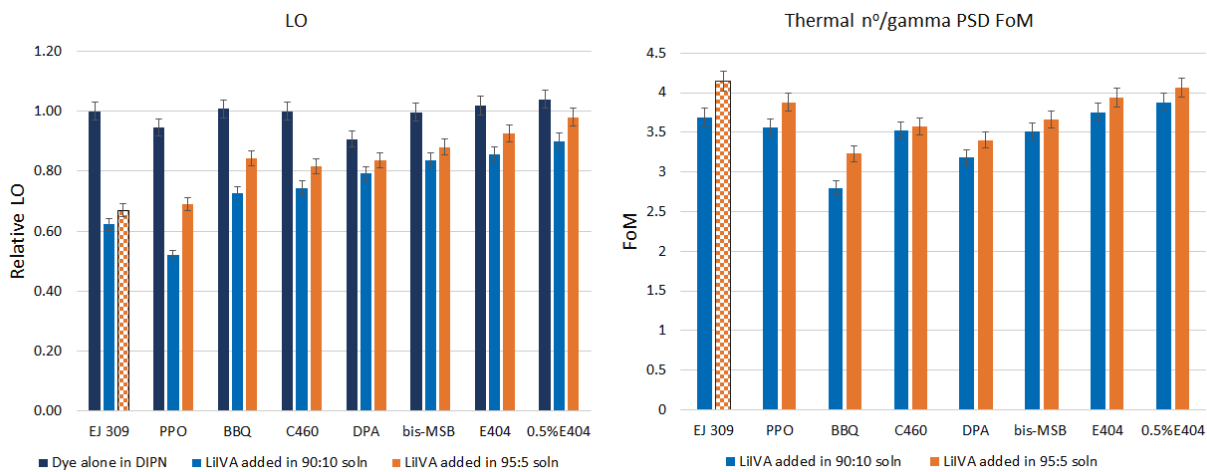

*Figure 5. Scintillation properties of unloaded EJ-309 and DIPN solutions obtained by the direct dissolution of 0.2 wt.% of individual dyes (unless shown differently for 0.5% of E404) in comparison to $^6$LiIVA-loaded solutions of the same composition containing 10% and 5% of IBA as a solubilizing agent: A – Results show that $^6$Li-loading produces the highest drop of the LO in PPO-containing EJ-309 and PPO-DIPN solutions; B – Respective thermal neutron/gamma PSD FoMs measured at the thermal neutron energy spot position for solutions containing 0.1 wt.% $^6$Li. Patterned bars correspond to an unstable solution that produces solid precipitation in several days after preparation. All other formulations found to be stable over more than one year of observations.*

Comparative analysis of LO and PSD (Figure 5B) also shows increase of FoM with decreasing concentration of acid in 95DIPN:5IBA solutions. However, while for most compositions there is a visible correlation between LO and PSD, the larger decrease of the LO in PPO-containing systems does not lead to a similar drop in PSD FoM. This mismatch may indicate that the absorptive species that reduce light output selectively quench the prompt light resulting from the deexcitation of singlet states. At the same time, these species may have a milder effect on the excited triplet states migration and annihilation that produce delayed light and PSD. While the current explanations of the effects are based on speculation that requires more detailed considerations, a practical outcome of the conducted studies is that the PPO-containing LSs may not be the best choice for preparation of the LSs loaded with $^6$Li in combination with solubilizing acids. Simpler processing and improved performance can be realized with the alternative to PPO solutions that, having milder reactivity with acids, may provide substantially higher LO than that obtained with PPO-containing formulations.

Broader option of different dyes offers another advantage related to solubility problems. For example, traditional bis-MSB used in EJ-309 and PROSPECT compositions has solubility less than 0.5 wt.% in most aromatic solvents. At very low concentrations, it may remain stable in LSs containing highly soluble PPO. However, additions of lithium salts may push its solubility to lower levels, making preparations of LSs with more than 0.1 wt.% of $^6$Li problematic. Use of more soluble single dyes, like E404, C460, and BBQ, which are soluble in DIPN at least at ~10 wt.%, create broader options that may allow production of LSs with higher $^6$Li loadings.

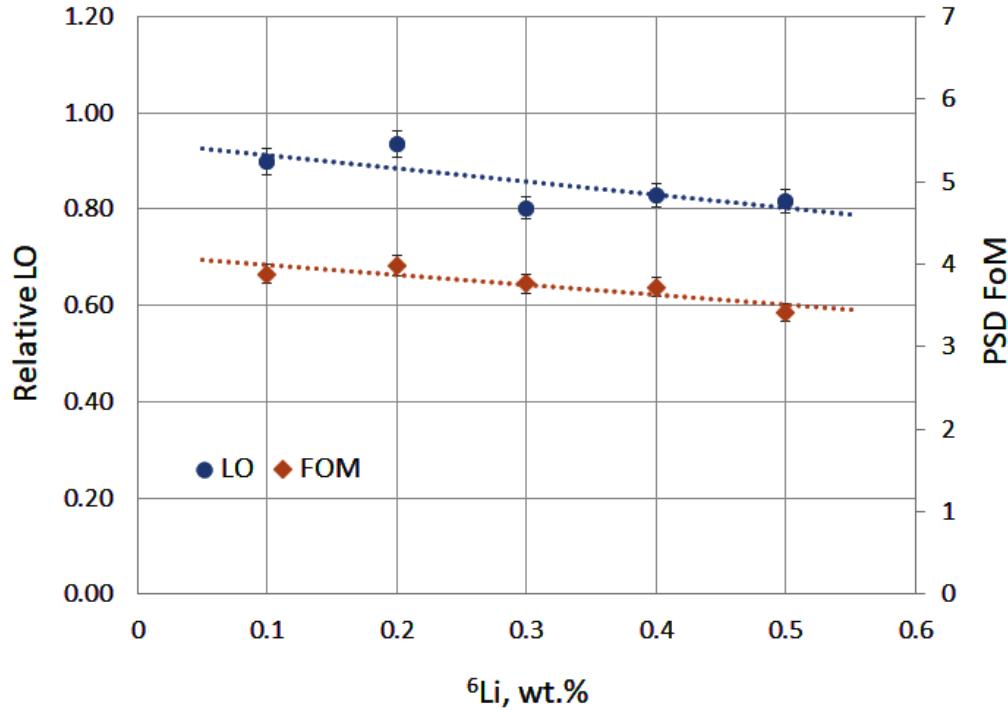

*Figure 6. Dependence of $^{137}$Cs LO and $^{252}$Cf PSD FOM on $^6$Li concentration in 90:10 DIPN:IBA solution containing 0.5% of E404 in combination with respective amounts of $^6$LiIVA. Solid precipitation was eventually observed at ≥ 0.4% $^6$Li, while lower concentration solutions remained stable during ~1 year of observations. A linear fit of these results has been added as a guide to the eye.*

Figure 6 presents results obtained with LSs containing 0.5% E404 dissolved in 90:10 DIPN:IBA solution made with different amounts of $^6$LiIVA corresponding to $^6$Li concentration up to 0.5 wt.%. Results show that there is practically no change in values of LO and PSD between 0.1 and 0.2 wt.% loads. Moreover, even at the maximum value of 0.5 wt.% $^6$Li, the total LO loss relative to unloaded EJ-309 does not exceed 20%, with still efficient PSD FoM of 2.5. It should be mentioned that precipitation was eventually observed at about 0.4 wt.%, while air-sealed solutions with 0.2-0.3 wt.% of $^6$Li remained stable during about a year of observations making these compositions potentially useful for applications requiring enhanced thermal neutron detection efficiency.

*3.1.4. $^6$Li as a part of a surfactant molecule*

An alternative approach to dissolution of $^6$Li would be to incorporate $^6$Li into a surfactant that forms a homogeneous solution in organic solvents. As described in the Experimental section, $^6$LiDOSS was synthesized in this work by replacing sodium by $^6$Li in the molecule of NaDOSS commercially produced as Aerosol AT, or AOT [22]. Like Tx-100, the AOT has a hydrophilic head and long hydrophobic tails that facilitate formation of micelles used in preparation of stable water-oil microemulsions. According to modeling and experimental studies [23], in the absence of water in non-polar systems, AOT can form "dry" micelles with a core containing Na, S, and O atoms surrounded by the dense aliphatic surface layers that make it highly soluble in aromatic solvents. Due to the similarity in properties of sodium and lithium, the formation of similar structures that screen the high polarity of the core was expected for the $^6$Li analog of AOT. Initial experiments confirmed the suitability for $^6$LiDOSS as an additive to LSs, since it was found to be easily dissolved in DIPN at high concentrations up to ~ 20 wt.% without the addition of any cosolvents like IBA.

Based on the molecular weight of the compound, obtaining 0.1 wt.% of $^6$Li requires a relatively large addition of 7.1 wt.% of $^6$LiDOSS. This addition produces less of a drop of LO or PSD compared to other compositions, likely due to the absence of solubilizing acid and/or the decreased polarity of $^6$Li in the molecule. As seen from results presented in Figure 7, addition of $^6$LiDOSS to single-dye solutions of bis-MSB, BBQ, or E404 leads to only 6-10% of LO drop, with preservation of the high PSD FoMs. However, the most remarkable improvement by using previous compositions, PPO was selectively affected by polar additions of acids and $^{6}Li$-salts. In the absence of such additions, $^{6}LiDOSS$-loaded EJ-309 also shows the highest FoM, largely determined by the PSD properties of PPO being less affected in this case by the $^{6}Li$ addition. It should be noted that the high solubility of $^{6}LiDOSS$ also allows for dissolution of larger concentrations of $^{6}Li$ (Figure 8). Since the solubility of the compound in DIPN is about 20%, loading to 0.3 wt.% $^{6}Li$ can be problematic, but solutions containing 0.2 wt.% of atomic lithium are stable to long-term precipitation. Like in the case of $^{6}LiIVA$ loading, higher concentration of $^{6}Li$ leads to a decrease of the LO. However, the downside of this new formulation is that at 0.2 wt.% $^{6}Li$, the LO drops about twice as much than the LO of a respective LS that contains $^{6}LiIVA$ (Figure 6). This larger drop may be attributed to a higher self-absorption, scattering, or dilution by the large non-aromatic fraction of $^{6}LiDOSS$ that at the increased scale may lead to larger LO loss and increasing attenuation.

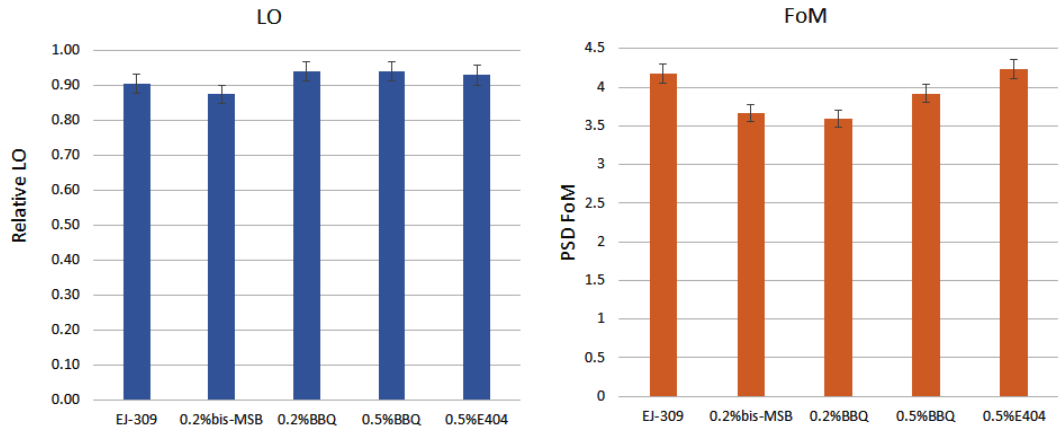

*Figure 7. LO and PSD measured with different LSs loaded with 7.1 wt.% of $^6$LIDOSS (0.1 wt.% of $^6$Li). Results show that $^6$Li-loaded LSs can be prepared with LO decrease not exceeding 6% compared to the unloaded EJ-309 standard. A remarkable property of $^6$LiDOSS is that, with only 10% of LO loss, it allows for the preparation of small-scale EJ-309-based LS with highest PSD among all tested compositions.*

$^6$LiDOSS is observed with EJ-309 that in this case loses only ~10% of the LO compared to 30-45% observed with other methods of EJ-309 loading of $^6$Li-loading. This observation is likely due to the absence of solubilizing acid and/or the low polarity of $^6$LiDOSS, confirming that in

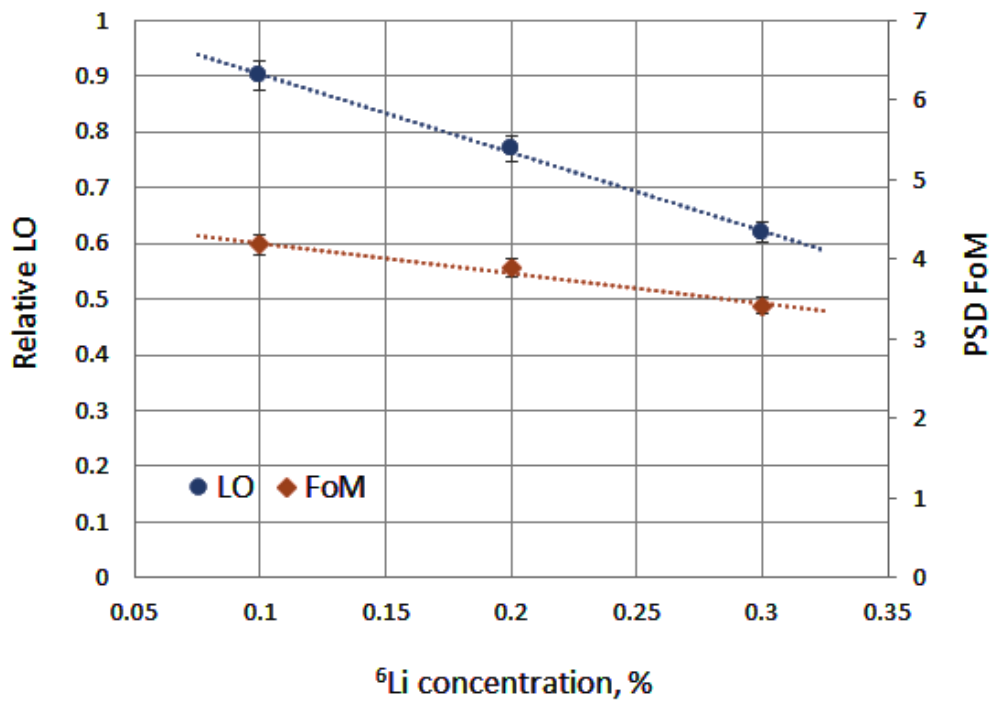

*Figure 8. Dependence of $^{137}Cs$ LO and $^{252}Cf$ PSD FoM on $^{6}Li$ concentration in EJ-309 solutions containing respective amounts of $^{6}LiDOSS$. Solutions proved to be stable to solid precipitation at ≤ 0.3% $^{6}Li$.*

## 4. Studies of different $^{6}Li$-loaded compositions at increasing scale

*4.1. Preliminary evaluation of attenuation*

Considering that potential applications of $^{6}Li$-loaded LSs may involve large-volume preparations, the most typical compositions measured in the small cuvettes (Ø2 cm x 1 cm) were additionally characterized with larger volume (Ø5 cm x 5 cm) quartz cuvettes, for an initial approximation of the attenuation that can become more pronounced at the increasing scintillator length. Figure 9 presents a compressed summary of the main formulations prepared and compared using two different volume samples. As expected, LO decreases in all tested samples at increasing size (Figure 9A). However, the degree of this decrease is different depending on the composition.

Analysis of results shows that loading of EJ-309 by $^6$Li-salts (Group I) may not be the best option for applications requiring high LO and low attenuation. This is the case even for the EJ-309 – $^6$LiDOSS formulation that, despite the high initial brightness, loses about 15% of its LO when the scintillator thickness is increased by a factor of 5. A similar difference in the LO between the samples of different scale is also observed in LSs prepared with the brightest single dyes dissolved in 90:10 DIPN:IBA solutions (Group II). Due to the high initial LO, samples of this group can be comparable to the EJ-309-$^6$LiDOSS composition. However, better initial performance and the smallest LO decrease measured with single-dye samples prepared in 95:5 DIPN:IBA solutions makes formulations of the Group III more attractive and promising for further studies.

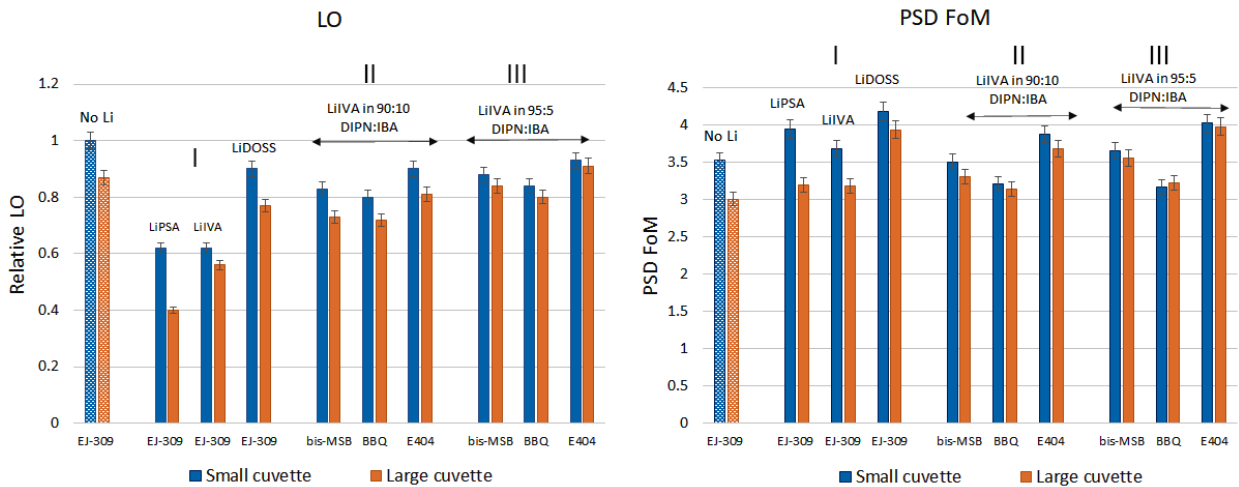

Figure 9. Results showing the difference of the $^{137}$Cs LO (A) and $^{252}$Cf PSD FOM (B) measured with the same LSs in different volume quartz cuvettes: Ø2cm x 1cm (small) and Ø5cm x 5cm (large). Unloaded EJ-309 sensitive only to fast neutrons (no Li) is given as a standard for a typical decrease of both parameters at increasing scale. The rest of composition are LSs containing 0.1 wt.% of $^6$Li added by different methods studied in the current work.

An interesting observation is that corresponding PSD of the tested samples (Figure 9B) does not necessarily directly correlate with the LO measured in samples of different scale. Despite a noticeable LO loss, $^6$Li-loaded EJ-309 (Group I) still preserves relatively high PSD (presumably determined by the exceptional discrimination properties of PPO). As for single-dye formulations loaded with $^6$LiIVA, this level of scaling does not produce any visible effect on PSD FoMs. The FoMs of these scintillators remain practically unchanged or only decrease by a small value (Groups II and III, Figure 9B).

*4.2. Large-scale measurements of selected formulations*

The overall results of Figure 9 showed that among EJ-309-based formulations, only loading with $^6$LiDOSS produces high LO comparable to that of the initial unloaded EJ-309 base. With the $^6$LiIVA-loading of large-scale LSs, higher LO and good attenuation properties could be expected in 95:5 DIPN:IBA solutions. To verify these results, the three best representatives of these groups (I and III) were tested for more detailed studies of LO, effective attenuation length, and PSD at a further increased scale (Table 2). The photo and the dimensions of acrylic cells used in experiments are presented in Figure 10A.

| Liquid | LO, PE/MeV | Attenuation length, cm | PSD FoM |
|---|---|---|---|
| EJ-309 | 706 ± 20 | 65.2 ± 1.8 | 1.86 ± 0.05 |
| EJ-309 + $^6$LiDOSS | 536 ± 15 | 57.5 ± 1.6 | 1.95 ± 0.05 |
| 0.2% bis-MSB + $^6$LiIVA | 621 ± 17 | 75.3 ± 2.1 | 1.68 ± 0.05 |
| 0.5%E404 + $^6$LiIVA | 718 ± 20 | 74.1 ± 2.1 | 2.03 ± 0.06 |

*Table 2. Compositions and respective parameters of tested $^6$Li-loaded formulations containing 0.1 wt. % of $^6$Li.*

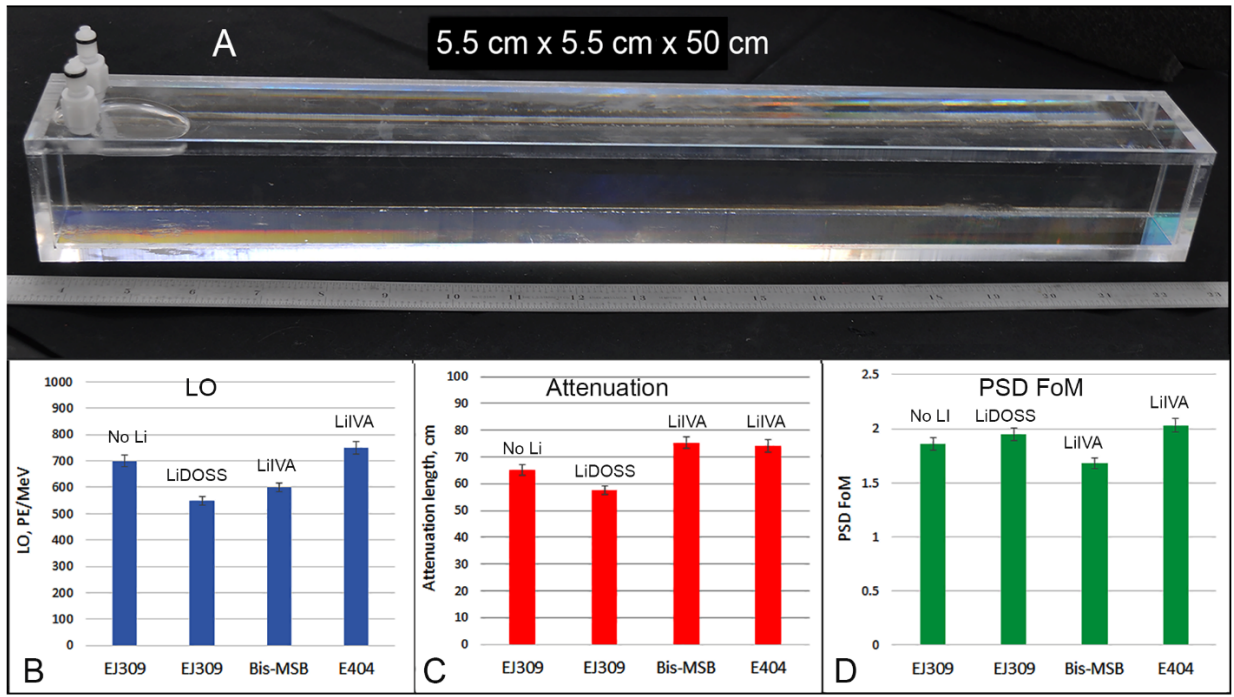

*Figure 10. A – Photo and dimensions of an acrylic cell used for large scale measurements; B – D – graphic representation of LO, effective attenuation length, and PSD FoM shown in Table 2. Similar to small-scale experiments, FoM for unloaded EJ-309 was calculated in the $^{137}$Cs CE energy range, while FoM of $^6$Li formulations correspond to the position of the thermal neutron spot. Bis-MSB and E404 formulations were prepared in 95:5 DIPN:IBA solutions.*

Analysis of the data (Figure 10 B-D) obtained with the large-scale samples indicates that among the three $^6$Li-containing formulations, EJ-309 loaded with $^6$LiDOSS, shows the highest decrease of the scintillation efficiency and the attenuation length compared to the EJ-309 test sample. The result is consistent with the observations made in smaller-scale experiments that suggested higher LO quenching by $^6$LiDOSS in comparison with $^6$LiIVA formulations (Sections 3.1.4). As shown in the previous section, despite these effects, PSD of this scintillator remains one of the highest (Figure 11) that, in combination with the easy preparation without additions of potentially corrosive acids, makes EJ-309:$^6$LiDOSS LS promising for certain large-scale applications requiring good neutron/gamma discrimination performance.

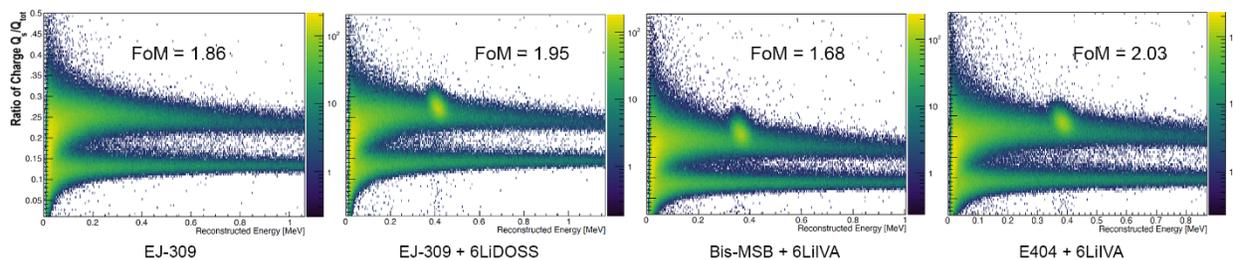

*Figure 11. Experimental PSD distributions obtained with unloaded EJ-309 standard and the large-scale $^6$Li-loaded liquids with compositions shown in Table 2.*

Results obtained with two other LSs, made with $^6$LiIVA added to single dye bis-MSB and E404 95:5 DIPN:IBI solutions are also consistent with the measurements of the preliminary studies. As expected, E404 + $^6$LiIVA composition shows the best values of all tested parameters that include the highest LO, attenuation lengths, and PSD FoM (Figures 10 and 11). For very large-volume applications, use of E404 can be problematic because of the high cost typical for all Exalite dyes. In this case, inexpensive and widely available bis-MSB or other similar dyes can be preferred as more acceptable alternatives that may still provide better LO and attenuation performance compared to the current formulations characterized by a 30-40% of LO loss relative to EJ-309.

5. **Conclusions**

The most important results obtained in this work are presented in Table 3 that shows that large volumes of $^6$Li-loaded LSs can be produced with LO or attenuation similar, or even exceeding those of the best unloaded LSs. Since both light output (measured in PE/MeV) or 'effective' attenuation are the quantities dependent on the specifics of the experimental setups, direct comparison of the absolute values to literature data or results obtained in different experiments is difficult. However, the performance of the studied formulations evaluated through a common test material, EJ-309, used in the current studies as a scintillation standard, clearly shows the advantages of the proposed methods that include simplicity of preparation combined with high performance and physical stability. Three-step scaling studies conducted in this work showed that selection of optimal compositions cannot be done based only on small-scale measurement results that can substantially change at larger dimensions. The methods also provide broader options for development of more advanced compositions that may combine high performance with acceptable cost. The additional benefit of new formulations can be much milder, or no corrosion effects produced by the low concentration of a week carboxylic acid in comparison to much more corrosive combination of HCl and water additions. And finally, it should be added here that the tested compositions by no means include all possible formulations that can be prepared with larger varieties of different dyes that, at increasing scale may offer better performance in future developments.

| Parameter/LS | EJ-309 | EJ-309+ [6]LiDOSS | Bis-MSB+[6]LiIVA | E404+[6]LiIVA | PROSPECT LS |
|---|---|---|---|---|---|
| LO | 1.0 | 0.76 ± 0.02 | 0.87 ± 0.02 | 1.02 ± 0.03 | 0.71 ± 0.02 [9] |
| Attenuation length | 1.0 | 0.88 ± 0.02 | 1.15 ± 0.03 | 1.15 ± 0.03 | n/a |

Table 3. Ratios showing fractions of LO and attenuation lengths relative to EJ-309 obtained in large-scale experiments with the three best $^6$Li-loaded formulations.

**Acknowledgements**

The work was performed under the auspices of the U.S. Department of Energy by Lawrence Livermore National Laboratory under Contract DE-AC52-07NA27344 and was supported by the LLNL-LDRD Program under Project No. 20-SI-003. LLNL-JRNL-846737